\documentclass[12pt,a4paper]{article}

\usepackage{amsmath}

\begin{document}
\title{On the invariance of Scaled Factorial Moments when 
original distribution is folded with the binomial}
\author{Marek Kirejczyk%
\thanks{On leave at: Physikalisches Institut, Universitaet Heidelberg,
Philosophenweg 12, D-69120~Heidelberg, Germany\@.} \thanks{%
email: Marek.Kirejczyk{@}fuw.edu.pl}\\ 
\normalsize Institute of Experimental Physics, Warsaw University\\
\normalsize Ho\.za 74, 00--714~Warszawa, Poland}
%}}
\date{}

%\begin{document}

\maketitle

\begin{abstract}
It is shown, that the Scaled Factorial Moments of any rank 
do not change if the original distribution is
folded with a binomial one.
\end{abstract}

Scaled Factorial Moments (SFMs) have been used for some time in the
analysis of event-by-event correlations and fluctuations, as they
allow for a ``direct'' access to the ``dynamical'' fluctuations of the
multiplicity~\cite{biaplo}. They were used (among others) for the
intermittency analysis~\cite{biaplo}, and recently they were also
suggested as a tool for testing the assumption of chemical
equilibration in the nuclear collision~\cite{refeq}.   For a selected
class of particles produced with low multiplicity --- ones that carry
produced and conserved in the reaction charge-like quantity -- the SFM
would be close to 1/2 if chemical equilibrium is reached.
  
One example of particles, that are good candidates for such an
analysis, are kaons when observed at SIS energies (beam kinetic energy up
to 2\,A\,GeV).   They carry positive strangeness, and as the beam
energies are close to the production threshold, their multiplicities
are small.
 
In this case the kaons are produced either as K$^+$ or K$^0$.  However
only one (usually charged) type is registered in the detector with
reasonable (but still smaller then one) probability.   This led to the
question --- how much the SFMs of the measured distributions differ
from the ``original''  SFMs? 
 
%For the sake of simplifying the formulae SFM of rank 2 is chosen for
%this discussion.   Repeating the reasoning for the other ranks is
%straightforward.

Factorial Moment of rank $j$ is defined as follows:
\begin{equation}
F_j= \langle I (I-1) \dots (I-j+1)  \rangle  = 
\sum_I^\infty I (I-1) \dots (I-j+1) P_I ,
\end{equation}
where $I$ is number of particles, and $P_I$ is the probability of
producing $I$ particles in an event.  $P_I$ has a normalized
distribution.

Scaled Factorial Moment of rank $j$ is then defined as
\begin{equation}
SF_j=\frac{F_j}{\langle I \rangle ^j}.
\end{equation}

Capital symbols, like $I$ and and $P_I$ denote the ``original''
distribution of produced kaons.   Now let us assume, that for each
produced kaon the probability of registering it is equal to $q$.  The
resulting, ``measured'' distribution will then be a fold of the
``original'' distribution with the binomial.   Denoting with small
symbols $i$ the number of registered kaons per event and and $p_i$
probability of an event with $i$ registered kaons one obtains:
\begin{equation}
p_i= \sum_{I=i}^\infty \binom{I}{i} q^i (1-q)^{I-i} P_I.
\end{equation}

The distribution of $p_i$ is normalized.  The following proof
illustrates the technique of reordering sums, which is used in the
later part of the paper.  One needs to note, that for any individual
case $i \leq I$, so each $I$ contributes only to terms with $i \leq I$.
\begin{equation}
\label{eq:norm_p}
\begin{split}
\sum_i^\infty p_i & = \sum_i^\infty \sum_{I=i}^\infty \binom{I}{i} q^i
(1-q)^{I-i} P_I 
 = \sum_{I=0}^\infty \sum_{i=0}^I \binom{I}{i} q^i (1-q)^{I-i} P_I \\
& = \sum_{I=0}^\infty P_I \sum_{i=0}^I \binom{I}{i} q^i (1-q)^{I-i} 
 = \sum_{I=0}^\infty P_I = 1.
\end{split}
\end{equation}
The normalization of $P_I$ and normalization of the binomial distribution
were used in eq.~\ref{eq:norm_p}.

Factorial Moment of the measured distribution ($f_j$) equals to:
\begin{equation}
\begin{split}
f_j & =\sum_i^\infty i (i-1) \ldots (i-j+1) p_i \\
& = \sum_i^\infty i (i-1) \ldots (i-j+1)
\sum_{I=i}^\infty \binom{I}{i} q^i (1-q)^{I-i} P_I.
\end{split}
\end{equation}

To calculate $f_j$ explicitly, the same reordering of the summation
as in eq.~(\ref{eq:norm_p}) is used.   One notes, that terms with $i < j$
are equal to zero and do not contribute to the sum, so in
reality the summation starts not with $i=0$, but with $i=j$.  At a point
$k$ is substituted for $i-j$ and $K$ for $I-j$.  Note, that
$(I-i)=(K-k)$.
\begin{equation}
\label{numerator}
\begin{split}
f_j & = \sum_{i=0}^\infty i (i-1) \ldots (i-j+1) \sum_{I=i}^\infty 
\binom{I}{i} q^i
(1-q)^{I-i} P_I = \\ &
      = \sum_{I=0}^\infty P_I \sum_{i=j}^I i (i-1) \ldots (i-j+1) \\
& \hspace{1em} \times \frac{I (I-1) \ldots (I-j+1) (I-j)!}{i (i-1)
\ldots (i-j+1) (i-j)! (I-i)!}q^j q^{i-j} (1-q)^{I-i} \\
     & = \sum_{I=0}^\infty P_I I (I-1)  \ldots (I-j+1) q^j 
\sum_{k=0}^K \frac{K!}{k! (K-k)!} q^k (1-q)^{K-k} \\
     & = q^j \sum_{I=0}^\infty P_I I (I-1)  \ldots (I-j+1) = q^j F_j.
\end{split}
\end{equation}

In order to calculate the scaling denominator one needs to 
calculate $\langle i \rangle $.  
%In formula~(\ref{eq:imean}) $k$ is substituted for
%$i-1$, $K$ for $I-1$ and $K-k$ for $I-i$
One notices, that $\langle i \rangle $ equals to $f_1$ (and $\langle I
\rangle  = F_1$), so one can use the results of
formula~(\ref{numerator}).
\begin{equation}
\label{eq:imean}
%\begin{split}
%\langle i \rangle  & = \sum_{i=0}^\infty i p_i = \sum_{i=0}^\infty i
%\sum_{I=i}^\infty \binom{I}{i} q^i (1-q)^{I-i} P_I \\
%     & = \sum_{I=0}^\infty P_I \sum_{i=0}^I i \frac{I (I-1)!}{i (i-1)! (I-i)!}
%q q^{i-1} (1-q)^{I-i}\\ 
%    & = \sum_{I=0}^\infty P_I I q \sum_{k=0}^K \frac{K!}{k! (K-k)!}
%q^k (1-q)^{K-k} 
%= q \sum_{I=0}^\infty I P_I = q \langle I \rangle .
%\end{split}
\langle i \rangle  = f_1 = q^1 F_1 = q \langle I \rangle 
\end{equation}

Finally the measured Scaled Factorial Moment, $sf_j$ appears
equal to the original one.
\begin{equation}
sf_j = \frac{f_j}{\langle i \rangle ^j} = 
\frac{q^j F_j}{(q \langle I \rangle )^j} =
\frac{q^j F_j}{q^j \langle I \rangle ^j} = 
\frac{F_j}{\langle I \rangle ^j} = SF_j.
\end{equation}


\begin{thebibliography}{9}
\bibitem{biaplo} A.~Bia{\l}as, R.~Peshanski, \textit{Nucl.\ Phys.}
\textbf{B 273}, 703 (1986); \textit{Nucl.\ Phys.} \textbf{B 308}, 857
(1988).
\bibitem{refeq} S.~Jeon, V.~Koch, K.~Redlich, X.\,-N.~Wang, 
\textit{Nucl.\ Phys.} \textbf{A 697}, 546 (2002).
\end{thebibliography}
\end{document}